\documentstyle[12pt,epsf]{article}

\def\beq{\begin{equation}}
\def\eeq{\end{equation}}
\def\beqa{\begin{eqnarray}}
\def\eeqa{\end{eqnarray}}

\begin{document}

\begin{flushright}
ITP-SB-98-43\\
NIKHEF/98-017\\
\end{flushright}

\vspace{3mm}
\begin{center}
{\Large\bf\sc Resummation of Threshold Corrections}\\
\vspace{5mm}
{\Large\bf\sc for Single-Particle Inclusive Cross Sections}
\end{center}
\vspace{2mm}
\begin{center}
Eric Laenen$^{1,2}$, Gianluca Oderda$^1$ and George Sterman$^1$\\
\vspace{7mm}
{\it $^1$Institute for Theoretical Physics\\
State University of New York at Stony Brook\\
Stony Brook, NY 11794-3840, USA} \\
\vspace{4mm}
{\it $^2$NIKHEF\\
P.O. Box 41882, 1009 DB Amsterdam, The Netherlands} \\
\vspace{6mm}
June, 1998
\end{center}

\vspace{7mm}
\begin{abstract}
We derive threshold resummations for
single-particle and single-jet inclusive cross sections, thus
generalizing previous results at fixed invariant mass to
a wider class of cross sections with
phenomenological interest.
We confirm the structure of our
resummed expressions by comparison with
explicit one-loop calculations for 
direct photons and heavy quarks.
\end{abstract}

\newpage

\section{Introduction}

A central success of quantum chromodynamics (QCD)
is the computation of inclusive,  
short-distance hadron-hadron cross sections.
This program combines
perturbative coefficient 
functions for partonic hard scattering with nonperturbative
parton distributions and fragmentation functions.
In another language, the 
nonperturbative matrix elements represent 
effective theories, associated with incoming or
observed hadrons, while the perturbative hard-scattering
functions match these theories to full QCD at
an adjustable scale, conventionally called the factorization scale.

The separation of short- and long-distance
dynamics in factorized cross sections is not
absolute, however, and soft gluon effects
persist in hard scattering functions.
Although infrared divergences 
cancel in perturbative coefficient functions
for hard scattering cross sections,
finite remainders can give substantial corrections at
higher orders of perturbation theory.  
The universality of soft gluon cancellation
makes it possible to resum these remainders to all orders.

These effects are potentially important for single-particle
inclusive cross sections for direct photons \cite{Diphoexp}
and heavy quarks \cite{QQexp} and for
very high-$p_T$ jet production \cite{sjetsexp}.  For the latter especially,
deviations from perturbative predictions may be a signal of
new physics.  Another outstanding question is the influence
of higher orders on global fits to parton distributions
\cite{OwensTung}.
For these applications, as well as for their intrinsic
interest, it is important
to examine higher orders in $\alpha_s$ and,
where possible, to develop
an extended quantitative formalism
to estimate their influence.  
In this paper, we present some of the
basic results necessary for this formalism.

The finite remainders of infrared cancellations
may be identified with
the regions in momentum space where the infrared singularities
arise.  Singularities
associated with ``partonic threshold",
at which the partons have just enough energy
to produce the final state that defines the
cross section, can influence 
the normalization and shape of the observed
cross section indirectly, through 
a buildup of logarithmically enhanced
singular distributions in higher order
corrections to hard-scattering functions.
Threshold enhancements of this sort
have been resummed to all orders for
the inclusive Drell-Yan cross section \cite{St87,CT} 
$d\sigma/dQ^2$,
at fixed pair invariant mass (PIM) $Q$, and more recently,
for pairs of heavy
quarks \cite{KS,BCMN} and jets \cite{KOS1,KOS2} at fixed invariant mass.  
However, many cross sections of phenomenological interest, both
for the detection of new physics and for the determination
of parton distributions, involve the
detection of single particles rather than pairs.
The purpose of this paper is to extend threshold
resummation to cross sections with single-particle 
inclusive (1PI) kinematics, including direct photon,
heavy quark and jet cross sections.  
The formalism and resummed 1PI cross sections are
discussed in the following section, and checked against
existing one-loop calculations in Sec.\ 3.

\section{Threshold Resummation}

A factorized single-particle 
(denoted collectively as $c$)
inclusive cross section at measured momentum $\ell$ 
may be written as
\beqa
E_\ell{d\sigma_{AB\rightarrow c(\ell)+X}\over d^3\ell}
&=&
{1\over S^2}\; \sum_{ab}\; \int dx\; dy\; \phi_{a/A}(x,\mu^2)\; \phi_{b/B}(y,\mu^2)\nonumber\\
&\ & \hspace{10mm} \times\, 
\omega_{ab\rightarrow c(\ell)+X}
\left( {s_4\over \mu^2},{t\over \mu^2},{u\over \mu^2},\alpha_s(\mu^2)\right)\, ,
\label{normalfact}
\eeqa
where for $c$ a photon or jet we introduce the kinematic
invariant $s_4$ by
\beq
s_4\equiv s+t+u\, ,
\label{sfourdef}
\eeq
in terms of partonic invariants
$s=(p_a+p_b)^2,\ t=(p_a-\ell)^2,\ u=(p_b-\ell)^2$.
With this definition, $s_4=M_X^2+\ell^2$ 
is the invariant mass squared of the QCD radiation 
recoiling against the observed particle or
jet plus the mass squared of the observed particle
or jet.  For the production of a pair of heavy
particles 
of mass $M$, the corresponding 
threshold quantity is found from
$s+t_1+u_1$, with $t_1=t-M^2$ and $u_1=u-M^2$.
In Eq.\ (\ref{normalfact}) we absorb into $\omega_{ab}$ ``fragmentation"
logarithms of $\ell^2/\mu^2$.  We shall not
treat these important corrections, or those associated with
photon isolation, here.
For jets, we can define $E_\ell=|\vec{\ell}|$, and
integrate over $\ell^2$ as part of the sum over final states.

Values of $x$ and $y$ for which $s_4$ vanishes 
define ``partonic threshold", at which the Born process
for direct photon production, or jet
production at fixed three-momentum $\vec{\ell}$,
uses all the available energy.
Integration down to $s_4=0$ leads to finite corrections
in the cross sections, due to singular distributions of 
the general form $\alpha_s^n\; [\ln^m(s_4/s)/s_4]_+$,
with $m\le 2n-1$.  It is the effects of such distributions
that threshold resummation organizes, to all orders in
perturbation theory.  It is easy to check that the steeper
the slopes of the parton distributions at values of $x$ and $y$ where
$s_4=0$, the larger are these effects.  In essence, threshold
resummation summarizes the interplay of parton luminosity
and the soft QCD bremstrahlung associated with the hard scattering.
The corresponding recoil of the hard scattering is ignored
for this purpose.  

Because the functions $\omega_{ab}$ are independent of the external hadrons,
they are computed in infrared-regulated perturbation theory
with $A$ and $B$ replaced by partons.
To organize  
singular distributions in $\omega_{ab}$ at threshold, we rely on 
further factorization
properties of the partonic cross section near $S_4=S+T+U=0$,
where we use capital letters for invariants defined
with respect to the overall process, $T_1=(p_A-\ell)^2$, etc.
Fig.\ 1 represents the factorization of the purely partonic
cross section $a+b\rightarrow c+X$ near threshold,
for direct photon production.  
For heavy quark and
jet production the corresponding factorizations
were discussed in Refs. \cite{KS,BCMN,KOS1,KOS2}.
The $h_{ab}$ absorb virtual parton propagators that are
off-shell by the order of the momentum transfer.
For convenience of notation, we define $H_{ab}\equiv h^*_{ab}h_{ab}$.
In direct photon production, the two lowest-order reactions
in $H_{ab}$ are $q+{\bar q}\rightarrow \gamma+q$ and $g+q\rightarrow \gamma+q$.
The functions $\psi$ incorporate the dynamics of
partons collinear to the incoming partons $a$ and $b$
($q\bar{q}$ or $gq$).  Up to corrections that are
finite at threshold, the $\psi$'s are flavor-diagonal.
The momenta of final state
particles associated with $\psi_{a/a}$ and $\psi_{b/b}$
may be approximated by $(1-x)p_a$ and $(1-y)p_b$, respectively.
The function $J^{(r)}$ represents partons recoiling against the photon,
with total momentum $p_R$.  Their
dynamics is summarized by a two-point function for the field of
flavor $r$ ($g$ or $q$).  Finally, the function $S(k_S)$ summarizes
the dynamics of soft gluons, of total
momentum $k_S$.
As indicated by the double lines in the figure, partons
involved in the hard scattering
are treated in the eikonal approximation in $S$ \cite{KOS1,KOS2}.  
For jet production, there is an additional jet
function associated with the collinear
particles that carry the observed momentum $\ell$.  
For jets and heavy quarks, $H\times S(k_S)$
is a product in the space of color exchange \cite{KS,KOS1,KOS2},
but for direct photon production $H$ and $S(k_S)$
are simply functions.  At threshold, the dynamics of each of the classes
of partons become independent, and the partonic cross
section reduces to a convolution \cite{KOS1}.
We now turn to the kinematics of QCD radiation
near threshold for 1PI processes, which plays a central role in resummation.   
\begin{figure}
\centerline{\epsffile{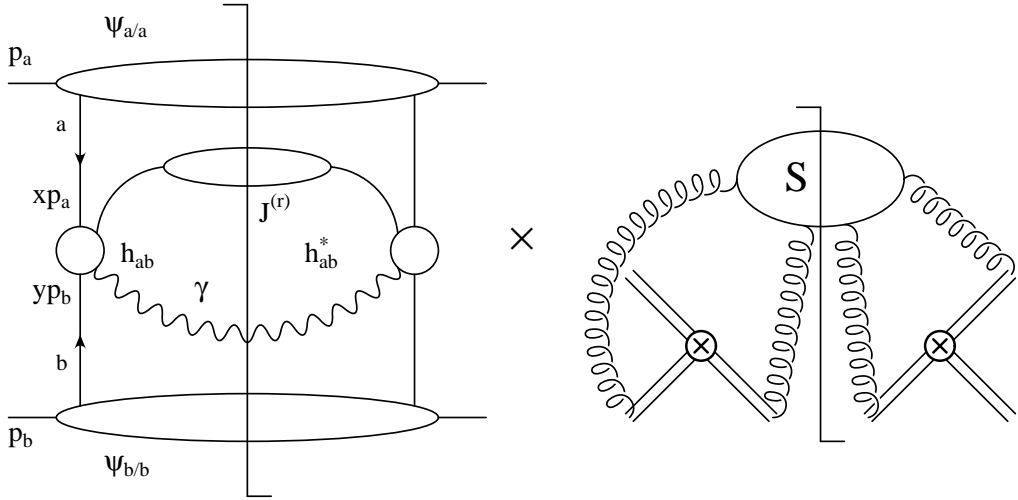}}
\caption[dum]{\small{
Factorization of direct photon production near threshold.
}}
\label{fig}
\end{figure}

Momentum conservation implies that $xp_a+yp_b=\ell+p_R+k_S$.
Near threshold, and neglecting corrections of order $S_4^2$,
we readily find that $S_4$ is a  sum of terms, each of which
may be associated with one of the functions identified above,
through the relations
\beqa
S_4 &=&  (1-x)2p_a\cdot \hat{p}_R+(1-y)2p_b\cdot \hat{p}_R
+2k_S\cdot \hat{p}_R+p_R^2+\ell^2\nonumber\\
&\equiv& \left[ w_a\left({u\over t+u}\right) + w_b\left({t\over t+u}\right) +w_S+w_R+w_\ell \right]\; S
\nonumber\\
&=& \left[(1-x)\left({u\over t+u}\right) + (1-y)\left({t\over t+u}\right)+{s_4\over S} \right]\; S\,  .
\label{sfourexpand}
\eeqa
The vector
$\hat{p}_R$ in the first line of Eq.\ (\ref{sfourexpand}) is the momentum of the 
recoiling jet (or heavy quark) at threshold.
In the center of mass frame, $\hat{p}_R^\mu=(\ell_0,-\vec{\ell})
\equiv \sqrt{S}\zeta^\mu$.  
In the second line, we introduce a set of dimensionless weights $w_i$, that
measure the contribution of each function in Fig.\ 1 to $S_4$.  
At threshold, each
of these weights vanishes.  The third line of Eq.\ (\ref{sfourexpand})
relates the overall $S_4$ to the corresponding quantity, $s_4$, 
defined in Eq.\ (\ref{sfourdef}), 
in the standard factorization, Eq.\
(\ref{normalfact}).
Note that $w_a\ne 1-x$ and $w_b\ne 1-y$, because these
variables refer to different choices of distributions: $w_a$ and $w_b$
to the functions $\psi$ of Fig.\ 1, and $x$ and $y$ to the distributions $\phi$
in Eq.\ (\ref{normalfact}).
In all ratios of kinematic factors, we use the quantities
$t$ and $u$ characteristic of the partonic hard scattering,
which is a good approximation at true threshold, when $x$ and $y$
approach unity.  

In these terms, the infrared-regulated, perturbative, and factorized
cross section, $a+b\rightarrow c+X$
at fixed $S_4$, may be written as an alternate convolution form \cite{CLS},
which directly reflects the organization of Fig.\ 1,
\beqa
E_\ell{d\sigma_{ab\rightarrow c(\ell)+X}\over d^3\ell}
&=&
H_{ab}(t,u)\; 
\int dw_a\; dw_b\; dw_S\; dw_R\; dw_\ell\nonumber\\
&\ & \hspace{-20mm} \times
\delta \left({S_4\over S} - w_a\left({u\over t+u}\right) - w_b\left({t\over t+u}\right) 
- \sum_{i=S,R,\ell}w_i \right)\; 
\nonumber\\
&\ & \hspace{-20mm} \times
\psi_{a/a}(w_a,p_a,\zeta,n)\; 
\psi_{b/b}(w_b,p_b,\zeta,n)\nonumber\\
&\ & \hspace{-20mm} \times J^{(c)}(w_\ell,\ell,\zeta,n)\; 
J^{(r)}(w_R,p_R,\zeta,n)\; S\left({w_S S\over\mu^2},\beta_i,\zeta,n\right)\, .
\label{threshconv}
\eeqa
When the observed particle $c$ is a photon, $J^{(c)}(\ell^2,\ell\cdot\zeta)$ 
may be replaced
by unity (neglecting fragmentation contributions),
and for $c$ a heavy quark both $J^{(c)}$ and $J^{(r)}$
may be absorbed into the hard-scattering function $H$.   
All the factors in Eq.\ (\ref{threshconv})
are evaluated in $n\cdot A=0$ gauge.
Essentially the same arguments for the factorized form Eq.\ (\ref{normalfact})
apply as well to
Eq.\ (\ref{threshconv}).  

In particular, the functions $\psi_{i/i}(w_i,p_i,\zeta,n)$
are the distributions of partons $i$ in parton $i$ with 
fixed values of momentum component $p_i\cdot\zeta$.  
They are constructed by direct analogy to ``center-of-mass"
distributions \cite{St87,KS,KOS1}, defined with
$\zeta^\mu=n^\mu=\delta_{\mu 0}$.
For example, as a matrix element in $n\cdot A=0$ gauge, $\psi_{q/q}$ is given by
\beq
\psi_{q/q}(w,p,\zeta,n)
=
{1\over 2N_c}\;
{p\cdot \zeta \over 2\pi p\cdot v}
\int_{-\infty}^\infty d\lambda\ {\rm e}^{-i(1-w)\lambda p\cdot\zeta}
\langle q(p)|\bar{q}(\lambda\zeta)\; {1\over 2}v\cdot \gamma\; q(0)|q(p)\rangle
\label{psiwdef}
\eeq
for an external quark of momentum $p$.  Here $v^\mu$ is the lightlike vector in the
direction opposite to $p^\mu$; for example, when  $\vec p$ is in the 3-direction,
$p\cdot v=p^+$.   The factor $(1/2N_c)$ reflects an average over spin and color.

Given any factorization of the form of Eq.\ (\ref{threshconv}), it was 
shown in Ref.\ \cite{CLS} that 
each of the factorized functions
satisfy evolution equations, whose solutions 
organize all logarithmic $S_4$-dependence at leading power.  
In addition, it is straightforward to verify that if
we choose the gauge vector $n^\mu$ such that $p_i\cdot n=p_i\cdot\zeta$ for
both $i=a,b$, the densities $\psi_{i/i}(w,p,\zeta,n)$
are equal to the center-of-mass densities ($n^\mu=\zeta^\mu=\delta_{\mu0}$),
at leading and next-to-leading
logarithm.  For direct photon, heavy quark and jet cross sections, such choices are
\beqa
n^\mu &=& {p_b\cdot\zeta\over p_a\cdot p_b}p_a^\mu
+
{p_a\cdot\zeta\over p_a\cdot p_b}p_b^\mu, \quad {\rm photon\ or\ jet\, ,}
\label{nphot}
\nonumber\\
n^\mu &=& \zeta^\mu, \quad\quad\quad {\rm heavy\ quark}\, .
\label{nhq}
\eeqa
For the heavy quark $\hat{p}_R$ is the momentum of the recoiling (unobserved) member of the pair.
In each case, $\psi_{a/a}$ is a function of 
$w_a$ and $p_a\cdot\zeta=-u/\sqrt{s}$ only, and $\psi_{b/b}$ a
function of $w_b$ and $p_b\cdot \zeta=-t/\sqrt{s}$ only.

We organize singular behavior at threshold in 
terms of a Laplace (or Mellin) transform,
\beq
{\tilde \omega}_{ab}\left( N,{t\over \mu^2},{u\over \mu^2},\alpha_s(\mu^2)\right)
=
\int_0^S {ds_4\over S}\; {\rm e}^{-N(s_4/S)}\,
\omega_{ab}\left( {s_4\over \mu^2},{t\over \mu^2},{u\over \mu^2},\alpha_s(\mu^2) \right)\, ,
\label{tilde omega}
\eeq
where $s_4$ is defined in  Eq.\ (\ref{sfourdef}) above.
In the transform, a singular distribution $[\ln^m(s_4/s)/s_4]_+$
produces $\ln^{m+1}N$, plus lower powers of $\ln N$. 
By comparing the moments with respect to $S_4$ of Eq.\ (\ref{normalfact}) 
for initial state {\it partons},
 $A=a$ and $B=b$ with moments of Eq.\ (\ref{threshconv}), 
and using the relation between $S_4$ and $s_4$ in Eq.\ (\ref{sfourexpand}), we derive \cite{KOS1}
\beqa
{\tilde \omega}_{ab}\left( N,{t\over \mu^2},{u\over \mu^2},\alpha_s(\mu^2)\right)
&=&
H_{ab}(t,u)\; 
\left[ {{\tilde\psi}_{a/a}(N{u\over t+u},p_a\cdot\zeta)\;
 {\tilde\psi}_{b/b}(N{t\over t+u},p_b\cdot\zeta)
\over {\tilde\phi}_{a/a}(N{u\over t+u},\mu^2)\; {\tilde\phi}_{b/b}(N{t\over t+u},\mu^2) }  \right]
\nonumber\\
&\ & \hspace{-25mm} \times  \tilde J^{(c)}(N,\ell\cdot\zeta)\;
{\tilde J}^{(r)}(N,p_R\cdot n)\; {\tilde S}\left({S\over N\mu^2},\beta_i,\zeta,n\right)\; 
+{\cal O}\left({1 \over N}\right)\, .
\label{threshmoment}
\eeqa
As in the case of $\omega_{ab}$, for each function $f$ the moment is $\tilde{f}(N)\equiv\int_0^1e^{-Nw}f(w)$.
For large $N$, the precise upper limit is unimportant.
The factors $t/(t+u)$ and $u/(t+u)$ are characteristic of
resummation for the single-particle cross section.
Solving the evolution equations for each of the functions in (\ref{threshmoment}) \cite{KOS1,CLS},
we derive an explicit expression for $\tilde{\omega}_{ab}(N)$, whose inverse transform \cite{AMS,LSvN,BC,CMNT} is
the fully-resummed hard scattering function for cross sections
with single-particle inclusive kinematics,
\beqa
{\tilde \omega}_{ab}\left( N,{t\over \mu^2},{u\over \mu^2},\alpha_s(\mu^2)\right)
&=&
\exp \Bigg \{ \sum_{i=a,b} E_{(i)}(N_i,p_i\cdot\zeta)\nonumber\\
&\ & \hspace{-20mm}  -\int_\mu^{p_i\cdot\zeta}
{d\mu'\over\mu'}\left[ \gamma_f(\mu')-\gamma_{ff}(N,\mu')\right]
\Bigg \}\; 
\exp \Bigg \{ \sum_{j=c,r} E'_{(j)}(N,p_j\cdot n) \Bigg \}
\nonumber \\
&\ & \hspace{-25mm} \times\; 
H_{ab}(t,u)\; \tilde{S}\left(1,\beta_i,\zeta\cdot n\right)\; 
\exp \Bigg \{  \int_\mu^{\sqrt{S/N}} {d\mu' \over \mu'}
2{\rm Re}\Gamma_S^{(ab)}(\mu')\Bigg\}\, .
\label{sigNHSfinal}
\eeqa
The first exponential, which gives the $N$-dependence of the
ratios of wave functions $\tilde{\psi}$ and $\tilde{\phi}$ in the $\overline{\rm MS}$ 
scheme, is precisely the same as for heavy quark and dijet production, 
\beqa
E_{(f)}\left(N_i,M\right)
&=&
-\int^1_0 dz \frac{z^{N_i-1}-1}{1-z}\; 
\Bigg \{\int^{1}_{(1-z)^2} \frac{d\lambda}{\lambda} 
A_{(f)}\left[\alpha_s(\lambda M^2)\right]\nonumber\\
&\ &   \hspace{5mm}
 +\frac{1}{2}\nu^{(f)}\left[\alpha_s((1-z)^2 M^2)\right]  \bigg \}\, ,
\label{omegaexp}
\eeqa
where again the extra factors in $N_a\equiv N(-u/s)$ and $N_b\equiv N(-t/s)$
reflect the kinematics of single particle inclusive cross section.
$A_{(f)}$ is given by the standard expression \cite{KT},
$A_{(f)}(\alpha_s) = C_f\left ( {\alpha_s/\pi} 
+(1/2) K \left({\alpha_s/ \pi}\right)^2\right )+\dots$,
where $C_f=C_F\ (C_A)$ for an incoming quark (gluon), and 
$K= C_A\; \left ( {67/ 18}-{\pi^2/ 6 }\right ) - {5/9}n_f$,
with $n_f$ the number of quark flavors.  Finally,
$\nu^{(f)}=2C_f\; ({\alpha_s/\pi})+\dots$.  
In the same exponential, the integral of the difference $\gamma_f-\gamma_{ff}$
gives the scale evolution of the ratio $\tilde{\psi}/\tilde{\phi}$
for flavor $f$ \cite{KOS1}.

The second exponential in Eq.\ (\ref{sigNHSfinal}) is associated with the final state jets.
For heavy quarks it is absent.
Adopting the notation of \cite{KOS1}, we have
\beq
E'_{(f)}\left(N,M\right)
=
\int^1_0 dz \frac{z^{N-1}-1}{1-z}\; 
\Bigg \{\int^{(1-z)}_{(1-z)^2} \frac{d\lambda}{\lambda} 
A_{(f)}\left[\alpha_s(\lambda M^2)\right]
+ B'_{(f)}\left[\alpha_s((1-z) M^2) \right] \bigg \}\, ,
\label{outexp}
\eeq
where $A_{(f)}$ is the same as in Eq.\ (\ref{omegaexp}), while the
flavor-dependent $B'$ is identified by comparing the
one-loop expansion of $E'$ to the one-loop two-point function
of the relevant parton.  For quarks and gluons the one-loop results are 
\beq
B'_{(g)} = {\alpha_s\over\pi}
\left\{ C_A\left[ {1\over 2}{n_f\over 3C_A}-{11\over 12}-1 +\ln (2\nu_g)\right]\right\}
\, ,\ 
B'_{(q)} = {\alpha_s\over\pi}\left\{C_F\left[-{7\over 4}+\ln (2\nu_q)\right ]\right\}\, ,
\label{bprimes}
\eeq
where we define $\nu_i\equiv(\beta_i\cdot n)^2/n^2$
for a particle of velocity $\beta_i$.  Finally, the 
last exponential in Eq.\ (\ref{sigNHSfinal}) is associated with soft
emission.  The ``soft anomalous dimension" $\Gamma_S$ 
depends on the kinematics of the hard scattering.
The one-loop matrix anomalous dimensions for jet production
were extensively discussed in Refs.\ \cite{KOS1,KOS2}, and for heavy
quarks in Ref.\ \cite{KS}.  In these cases,
the exponentials of the matrix soft anomalous dimension are
ordered, and occur in traces with matrices of hard scattering
functions.

We note that exponents from both the incoming and outgoing jets
are double logarithmic.  For the initial-state
jets they are positive, and enhance the cross section, but for the recoiling jet 
they are negative, and suppress it.
They are already present in the 
singularities at partonic threshold in the explicit
one-loop calculation of direct photon production,
as we will see in the next section.

The suppression associated
with the recoiling final state jet 
for direct photon production tends to oppose the enhancement
that is found from initial state jets in the production of 
heavy pairs \cite{St87,CT,LSvN,BC,CMNT}.
As pointed out in Ref.\ \cite{KOS1},
however, this relative suppression depends on the
manner in which the cross section is constructed.  The distinguishing 
criterion is whether
the cross section is defined in such a  way that partonic
threshold requires that $\ell^2=p_R^2=0$.  For a jet
or photon at fixed 3-momentum this is indeed the case, as we
see in Eq.\ (\ref{sfourexpand}).  Even a slight smearing of the jet momentum, 
 however, such as
in the cross section $d^2\sigma_{\rm jet}/dT\: dU$,
allows $p_R^2$ and $\ell^2$ to vary, and eliminates
double-logarithmic suppression due to final state interactions.

\section{One-loop Expansions}

We now verify that
the one-loop expansion of Eq.\ (\ref{threshconv}),
and therefore  Eq.\ (\ref{sigNHSfinal}),
indeed reproduces the singular functions for direct photon production
given in Ref.\ \cite{Dgamma1,GordonVogel}, and similarly that, in the case of heavy quark production,
the expansion of Eq.\ (\ref{sigNHSfinal}) 
reproduces the one-loop singular threshold behavior 
given in \cite{Mengetal}.
Note that to one loop the individual
contributions simply add up in Eq.\ (\ref{threshconv}).
In addition, for finite contributions in the $\overline{\rm MS}$ 
scheme, we may take $x=y=1$, because at zeroth order
$\phi_{a/a}(x)=\delta(1-x)$, and $\phi_{b/b}(y)=\delta(1-y)$.
Therefore, by the third line in Eq.\ (\ref{sfourexpand}),
$S_4=s_4$, and  we need not distinguish between 
these two variables at one loop.  It is important
to keep in mind that, following our comments after Eq.\ (\ref{sfourexpand}),
 $x=1$, $y=1$ does {\it not} imply $w_a=0$, $w_b=0$.
The value of the calculations below, of course, is not
to rederive known results, but to confirm the
exponentiated forms that organize singular distributions to all orders.

\subsection{Direct Photon Production}

The two lowest order partonic subprocesses for direct photon
cross sections are 
the ``Compton" process, $q_a(p_a)+g(p_b)\rightarrow q_r(p_r)+\gamma(\ell)$
and the ``annihilation'' process, 
$q_a(p_a)+{\bar q}_b(p_b)\rightarrow g(p_r)+\gamma(\ell)$.
For ease of comparison we cast our answers in terms of the kinematic variables
used by Gordon and Vogelsang (GV) in Ref.~\cite{GordonVogel}:
\beq
v \equiv {s+t \over s}\,,\;\; z \equiv {-u \over s+t}\,,
\eeq
so that $1-z = s_4/(s+t)$ (to avoid confusion,
we use $z$ rather than GV's $w$).
The Born cross sections for these two subprocesses
are given by 
\beqa
v(1-v)z\frac{d^2\sigma_{q\bar{q}}^{(0)}}{dv\, dz} & = & {2 C_F \over N_c}
\frac{\pi \alpha \alpha_s e_q^2}{s}  T_{q\bar{q}}\delta(1-z) \,,
\label{born1}\\
v(1-v)z\frac{d^2\sigma_{qg}^{(0)}}{dv\, dz} & = & {1 \over N_c}
\frac{\pi \alpha \alpha_s e_q^2}{s} T_{qg} v  \delta(1-z) \,,
\label{born2}
\eeqa
with
\beq
T_{qg} = 1 + (1-v)^2 \;,\;\; T_{q\bar{q}} = v^2 + (1-v)^2\, .
\eeq
In the above,
$N_c=3$ is the number of colors, and $e_q$ the electric 
charge of the quark.
The contributions from the singular functions in the 
NLO corrections found by GV may be written as
\beq
v(1-v)z s \frac{d^2\sigma_{ij}^{(1)}}{dv\, dz}  =  
\alpha \alpha^2_s e_q^2 \left\{ c_3^{ij} \Bigg[\frac{\ln(1-z)}{1-z}\Bigg]_+
+c_2^{ij} \Bigg[\frac{1}{1-z}\Bigg]_+
+c_b^{ij} \Bigg[\frac{1}{1-z}\Bigg]_+ \ln{\mu^2\over s}\right\}\,,
\eeq
where $ij = qg$ or $q\bar{q}$ and $\mu$ is the factorization scale. 
To derive the $c^{ij}$ in our formalism
we need the one-loop expressions for the exponents in 
Eq.~(\ref{sigNHSfinal}), or, equivalently, the one-loop
corrections to the functions in Eq.~(\ref{threshconv}).
Both direct photon production subprocesses receive contributions from 
two incoming and one outgoing jet function, and from 
the relevant soft function.

As stated in the previous section, one may define all functions
in Eq.~(\ref{threshconv}) as operator matrix elements \cite{KS,KOS1,St87}. 
We assume they are all normalized to $\delta(w_i)$, where the weights
$w_i$ are defined in Eq.~(\ref{sfourexpand}).
We have computed the one-loop contributions to these matrix elements
in $n\cdot A=0$ gauge, with $n^\mu$ chosen as in 
Eq.~(\ref{nphot}). 

The one-loop incoming-parton contribution 
in the $\overline{\rm MS}$ scheme is given by
\beq
\psi_{a/a}^{(1)}(w,n)= 2 C_a \left[{\ln w \over w}\right]_+ -
C_a {1\over w_+} + C_a {1\over w_+} \ln(2 \nu_a )
-C_a \ln\left({\mu^2\over s}\right){1\over w_+}\,,
\label{psioneloop}
\eeq
where $a=q,g$, with $C_q = C_F$ and $C_g = C_A$,
and where we recall that $\nu_i\equiv(\beta_i\cdot n)^2/n^2$
for a particle of type $i$ and velocity $\beta_i$.

Correspondingly,
the correction to this order for the outgoing gluon jet is (see Eqs.~(\ref{outexp})
and (\ref{bprimes}))
\beq
J^{(g),(1)}(w,n) = -C_A \left[{\ln w \over w}\right]_+
+C_A\left({1\over 2}{n_f\over 3 C_A} - {11\over 12} - 1 + 
\ln(2\nu_g)\right){1\over w_+}\,,
\eeq
and for the outgoing quark jet \cite{St87}
\beq
J^{(q),(1)}(w,n) = -C_F\left[{\ln w \over w}\right]_+
+C_F\left(- {7\over 4} + \ln(2\nu_q)
   \right){1\over w_+}\, .
\eeq
Notice that, indeed, the signs of the double-logarithmic 
terms in the $\psi$'s and
the $J$'s correspond to Sudakov enhancement and supression,
respectively.

The soft functions $S$ in  Eq.~(\ref{threshconv}) are ``eikonal"
cross sections constructed from  Wilson lines,
path-ordered exponentials of the gauge fields, in
color representations and along paths that
reflect the incoming and outgoing partons at threshold
\cite{KS,KOS2}.  In the case of direct photon production,
there are two Wilson lines in the fundamental representation,
representing quarks (antiquarks),
and one in the adjoint, representing the gluon,
coupled at a vertex $T_a^{(F)}$,
the generator of $SU(N_c)$ in the quark representation.   
At one loop,  soft functions are of the form
 $(1/w_+)2{\rm Re}\,\Gamma_S^{(ij)} $, where $\Gamma_S^{(ij)}$ is 
the one-loop anomalous dimension of the vertex, labelled by 
the equivalent initial state partons $i$ and $j$.

The rules necessary for the computation of the $\Gamma_S^{(ij)}$
are given in Refs.\ \cite{KS,KOS2}.  
Straightforward calculation shows that 
\beq
\Gamma_{S}^{(qg)} = \frac{\alpha_{s}}{2\pi}
\left\{C_A\left[-\ln \left(\frac{2u}{t}\right)+1-i\pi-\ln(\nu_g) \right]
+C_F\left[2\ln\left(\frac{-u}{2s}\right)+2-\ln(\nu_{q_a}\nu_{q_r})\right]\right\}, 
\label{eq:Aanofi}
\eeq
for the Compton process, and
\beqa
\Gamma_S^{(q{\bar q)}} &=& 
\frac{\alpha_{s}}{2\pi}\Bigg\{C_A
\left[\ln\left(\frac{tu}{2s^2}\right)+1+i\pi-\ln(\nu_g)\right]
\nonumber\\
&\ & \quad\quad 
+C_F\left[-2\ln(2)+2-2i\pi-\ln(\nu_{q_a}\nu_{\bar{q}_b})\right]\Bigg\}\,,
\label{eq:Banofi}
\eeqa
for the annihilation process.
Substituting the above results into  Eq.~(\ref{threshconv}),
accounting carefully for the kinematic factors in the
delta function that relates all the weights,
and multiplying with the Born cross sections, given in 
Eqs.~(\ref{born1}) and (\ref{born2}), gauge dependence cancels, and we find
\beqa
c_3^{qg} & = & T_{qg}{1\over N_c} (C_F+2N_c) v, \;\; c_b^{qg} = - T_{qg}{1\over N_c}(C_F+ N_c)v, \;
 \nonumber \\ 
c_2^{qg} &= &T_{qg}\left[ -\ln\left({1-v\over v}\right) 
   - C_F\Bigg({3\over 4N_c} -  {1\over N_c}\ln v \Bigg)\right]v \,,\nonumber \\
c_3^{q\bar{q}}  &=&  T_{q\bar q}{2 C_F \over N_c} (4C_F-N_c), \; c_b^{q\bar{q}}  
=  - T_{q\bar q}{4C_F^2\over N_c} 
\nonumber\\
c_2^{q\bar{q}}& =& T_{q\bar q}\left[{C_F n_f \over 3 N_c} - {11\over 6} C_F + 2 C_F \ln(1-v) 
        -4 {C_F^2\over N_c} \ln\left({1-v\over v}\right)\right]\, .
\eeqa
These coefficients agree precisely with those
of GV in Ref.~\cite{GordonVogel}.  We note that the coefficients of
the leading terms exhibit contributions from final state jets, which
act to suppress the cross section, as discussed at the end of
Sec.\ 2.  From the initial state functions $\psi$ alone, these
coefficients would have been $2C_F+2C_A$ for the Compton process, 
and $4C_F$ for the annihilation process (the same as in the
Drell-Yan cross section).  

\subsection{Heavy Quark Production}

As a further illustration, we consider heavy quark
production through the partonic subprocess 
$q(p_a)+\bar{q}(p_b)\rightarrow \bar Q(l)+X$.
The Mandelstam variables $s,t_1,u_1$ are defined below
Eq.~(\ref{sfourdef}), where now $s_4 = s+t_1+u_1$.
We shall derive the one-loop singular functions
in this process, given in Ref.~\cite{Mengetal},
in the $\overline{\rm MS}$ scheme.
The results may be represented as 
\beq
s^2 \frac{d^2\sigma_{q\bar{q}}^{(1)}}{dt_1\, du_1}  =  
{\alpha_s\over \pi} \sigma^{(0)} 
\left\{ c_3 \Bigg[\frac{\ln(s_4/m^2)}{s_4}\Bigg]_+
+c_2 \Bigg[\frac{1}{s_4}\Bigg]_+
+c_b \Bigg[\frac{1}{s_4}\Bigg]_+ \ln{\mu^2\over m^2}\right\}
\eeq
with $\mu$ the factorization scale.  The plus distribution in terms
of the dimensionful variable $s_4$ may be represented as
\beq
\Big[{\ln^{i}(s_4/m^2)\over s_4}\Big]_+
= \lim_{\Delta \rightarrow 0} \Bigg\{
{\ln^{i}(s_4/m^2)\over s_4} \theta(s_4\ -\Delta)
+ \frac{1}{i+1}\ln^{i+1}\Big({\Delta\over m^2}\Big)\, \delta({s_4}) \Bigg\}\, .
\label{s4distdef}
\eeq
The function $\sigma^{(0)}$ is the lowest order
cross section (see Ref.~\cite{Mengetal})
with the factor $\delta(s_4)$ removed.

We follow the same methods as in the previous subsection, 
and use the one-loop results for the $\psi_{i/i}$
densities of Eq.~(\ref{psioneloop}).
For the case at hand we find from Eq.~(\ref{sfourexpand})
\beq
w_a = {s_4\over m^2}\Bigg({m^2\over -u_1}\Bigg)\, ,\
w_b = {s_4\over m^2}\Bigg({m^2\over -t_1}\Bigg)\, .
\eeq
Because the heavy quark mass prevents collinear singularities,
all the final state contributions can be included in the soft function,
via its anomalous dimension. Therefore we can replace both
final state jet functions in Eq.~(\ref{threshconv}) by unity.
(Equivalently, we absorb their finite corrections into the
hard-scattering function.)
Although the soft anomalous dimension is a
matrix in the space of color tensors, the Born cross section in
the $q\bar{q}$ channel projects out only the octet-octet component. 
In Ref.~\cite{KS} this was computed to be\footnote{Note that
we have not normalized $\Gamma_S$ to the Drell-Yan
soft anomalous dimension, so that the expression
(\ref{gammashq}) has an extra term $+\alpha_s C_F/\pi$
compared to the result in \cite{KS}.}
\beqa
\Gamma_S^{{\bf 8}} & = & {\alpha_s\over \pi}
\Big\{ C_F\Big[ 4\ln\Big({u_1\over t_1}\Big) - 
 \ln(2\sqrt{\nu_q \nu_{\bar{q}}}) - L_\beta - \pi i\Big]\nonumber\\
&\ & + {C_A \over 2}\Big[-3 \ln\Big({u_1\over t_1}\Big)
-\ln\Big({m^2 s \over u_1 t_1}\Big) + L_\beta + \pi i\Big]\Big\}\, ,
\label{gammashq}
\eeqa
where
\beq
L_\beta=\frac{1-2\,m^2/s}{\beta}
\left\{ \ln\left( \frac{1-\beta}{1+\beta} \right) + {\rm{i}}\pi \right\},\;\;
\beta = \sqrt{1-4m^2/s}\, .
\eeq
Its one-loop contribution is again $(1/w_+)2{\rm Re}\,\Gamma_S $.
Combining these results, we find the same $\overline{\rm MS}$
scheme singular functions 
for heavy quark production in the $q\bar{q}$ channel as in 
Eq.~(28) of Ref.~\cite{Mengetal}:
\beqa
c_3 & = & 4 C_F  \nonumber \\
c_2 &= & C_F\Big[ -2\ln\Big({-u_1\over m^2}\Big)-2\ln\Big({-t_1\over m^2}\Big)
 -2 + 2 \ln\Big({s\over m^2}\Big) \nonumber \\
&\ & - 8\ln\Big({u_1\over t_1}\Big) -2{\rm Re}L_\beta \Big]
+C_A\Big[-3\ln\Big({u_1\over t_1}\Big) 
 - \ln\Big({m^2 s \over u_1 t_1}\Big) + {\rm Re}L_\beta\Big] \nonumber \\
c_b & = & - 2 C_F\, .
\eeqa
Note that in order to obtain the resummed heavy quark production
cross section in the DIS factorization scheme, one would
divide Eq.~(\ref{threshmoment}) by 
$( \tilde{F}_{2,a}/\tilde{\phi}_{a/a})\times
(\tilde{F}_{2,b}/\tilde{\phi}_{b/b})$ with 
$F_{2,a}$ the hard part of the deep-inelastic scattering
process \cite{CLS}.

Finally, we would like to make a general remark on the
connection of resummed cross section formulas for
single-particle inclusive (1PI) and 
pair-invariant mass (PIM) kinematics.
As  observed in Sec.\ 2, 
to next-to-leading logarithm, for $p_i\cdot n = p_i\cdot \zeta$, 
the densities $\psi_{i/i},\;i=q,g$ are all identical, 
whether they fix the energy or some other component of the incoming
parton momentum.
The relation of the weight $w_i$ to the total 
weight $s_4$ is in general different for 1PI kinematics
(Eq.~(\ref{sfourexpand})) and PIM kinematics 
\cite{St87,KOS1}.
Therefore the only extra terms in the contributions from the 
incoming partons, after transforming $\zeta^\mu$ (and therefore
$n^\mu$) from $\delta_{\mu 0}$
to $\hat{p}_R^\mu$, arise from the weight relation 
in Eq.~(\ref{sfourexpand}).
Performing the same transformation in the soft function in 
Eq.~(\ref{threshconv}) again yields extra terms, but 
here they are due to the
gauge dependence of the soft anomalous dimension. 
Kinematic effects from the weight relation are 
absent. 

The above observations apply when the vector $\zeta^\mu$ is time-like, 
for both PIM and 1PI kinematics.
In this manner one may transform the PIM resummed cross section
into the 1PI resummed cross section
(\ref{sigNHSfinal}), and back, the 
differences being easily computed by changing the gauge in the
soft anomalous dimension $\Gamma_S$, and
by replacing the moment variable $N$
by $N_a$, for incoming parton $a$, as in Eq.~(\ref{omegaexp}).

\section{Conclusion}

We have already mentioned the relevance of 
resummed single-particle cross sections to the theory and phenomenology
of QCD at high momentum transfer.  In the context of single-particle
inclusive cross sections, we have explored how threshold resummations 
provide information on higher-order
corrections.  As in the cases of heavy quark pair and Drell-Yan cross
sections, the resummed exponents $E$ and $E'$ for threshold resummation
contain infrared renormalons,  which  must be eliminated by a prescription
that depends, in  general, on nonperturbative paramaters.  In certain
cross sections at high scales, cross sections  may be quite independent
of these parameters, but this subject merits  further study.
We believe that the formalisms outlined here will find useful applications \cite{LaeMoch}.

Before concluding, we should point out that 
threshold resummation is not the only
possible organization of higher-order corrections
associated with soft gluon emission. 
 Of particular interest for the kinematic shape 
of 1PI cross sections
is the resummation of enhancements associated with
points in phase space at which partonic transverse momenta
vanish  \cite{Hustonetal,LaiLi}.  This would lead to
a ``$k_T$-resummation" for 1PI cross sections.  Such
a formalism exists for
the Drell-Yan and related processes \cite{CS81,CSSDY,DYgs} at measured 
(small) pair $Q_T$, but we know of no fully-developed method
for $k_T$-resummation in single-particle kinematics, or discussion of 
its relation to threshold resummation.  
 We believe that the organization of threshold
singular distributions will also be a valuable step toward
a fuller control over higher-order corrections including
transverse momentum effects.

\subsection*{Acknowledgements}

We thank Nick Kidonakis, Sven Moch and Jack Smith for many conversations
and insights.  E.L.\ wishes
to thank the Institute for Theoretical Physics, Stony Brook
for its hospitality during much of this project.
This work was supported in part by the National Science Foundation,
under grant PHY9722101.

\end{document}